\documentclass[
superscriptaddress,
groupedaddress,
prl,
amsmath,
amssymb,
twocolumn,
]{revtex4}

\usepackage{graphicx}  
\usepackage{dcolumn}   
\usepackage{bm}        
\usepackage{amssymb}   

\begin{document}



\title{Photonic Phase Transitions in Certain Disordered Media}
\author{Chin Wang}
\affiliation{Department of Physics, Beijing Institute of Technology, Beijing, China 100081}
\affiliation{Department of Physics, Hunan University of Arts and Science, Changde, China 415000}
\author{Luogen Deng}
\affiliation{Department of Physics, Beijing Institute of Technology, Beijing, China 100081}

\date{\today}

\begin{abstract}
In this paper we develop a generalized mode-expansion scheme for the vector lightwaves propagating in 3D disordered media,
and find the photonic phase transition from an extended-state (ES) phase to a mixed-state (MS) phase as a localization intensity parameter $G$ rises above $1/\sqrt{2}$. For the disordered media consisting of plenty of uniform spherical scattering particles, we formulate this phenomena at first in terms of actual optical and geometric parameters of the disordered media.
\end{abstract}

\maketitle


In spite of the fascinating behavior of the classical phase transitions many researchers also devoted their great enthusiasm to some novel phase transitions concerning with light or photons~\cite{Wiersma1997,Pier2000,Greentree2006,Novoa2010}.
Recently, structural strain in photonic resonator crystals has been used to induce a phase transition between different light modes~\cite{Pier2000},
a photonic system has been designed that may undergo a Mott insulator to superfluid quantum phase transition~\cite{Greentree2006}, and
certain common media, such as air or oxygen, have been proposed to support the propagation of steady light waves that appearing in the fermionic and liquid phases, and the transition from the former to the latter~\cite{Novoa2010}.
Various phase transitions of light are essentially the results of delicate light-matter interaction.
A familiar example is that the localization-delocalization transition of light modes in moderately disordered dielectric superlattices is sensitively dependent on the dielectric mismatch between scattering structures and background~\cite{John1987}.
However, for the more general case of fully disordered dielectric media, the perturbation technologies based on the band theory are inapplicable due to the absence of the lattice symmetry in these dielectric structures.
Likewise, tracing the process of the multiple scattering of light in random scattering media may not be a wisdom, as in this scheme one had to deal with the excessively complicated mechanism of the dependent scattering of light, and consequently the transport mean free path and the wavelength of photons need to be renormalized.

The study reported in this paper has been motivated by the difficulties mentioned above.
We consider that in ordinary position space disordered dielectric media have complex representations of refractive index profiles while usually in wave vector space sparse components, and it is an interesting issue weather some information for light localization can be distilled directly from the optical structure factors (OSFs) of disordered dielectric media, which are defined by the Fourier transformation of optical potentials.
For the idealized linear, but inhomogeneous, dielectric media we present the eigen equation of light waves by expanding both light fields and optical potentials in any set of generalized Fourier basis.
Based on this equation we derive a necessary condition supporting the strong localization, namely, the Anderson localization~\cite{Anderson1958} of light, and predict the photonic phase transitions between the ES phase and the MS phase.
For the disordered media consisting of plenty of uniform spherical scattering particles, we formulate this phenomena at first, to our best knowledge,  in terms of actual optical and geometric parameters of the disordered media.

We start from the wave equation for the complex electric field $\textbf{E}$ of radiation in idealized linear and isotropic polarized, but inhomogeneous, dielectric media,
\begin{equation}\label{complex equation}
\nabla ^2 {\bf{E}}  - \mu \frac{{\partial \sigma{\bf{E}}}}{{\partial t}} - \mu \frac{{\partial ^2 \varepsilon{\bf{E}}}}{{\partial t^2 }}-\nabla \left( {\nabla  \cdot {\bf{E}}} \right) =0~,
\end{equation}
where $\varepsilon$, $\mu$ and $\sigma$ are, in general, the scalar position- and frequency-dependent dielectric constant, magnetic permeability and conductance, respectively. Below we will consider media for which the magnetic permeability is approximately constant.
The single-frequency solution of the Eq.(\ref{complex equation}) is assumed to have an expression of $\textbf{E}(\textbf{r},t) = {\bm \Psi}(\textbf{r})e^{ - i\omega t}$,
and consequently the electric amplitude ${\bm \Psi}(\textbf{r})$ fulfills
\begin{eqnarray}
\label{wave equation2}
\nabla \times\left( {\nabla \times {\bf{\Psi }}({\bf{r}})} \right) - U(\omega ,{\bf{r}}){\bf{\Psi }}({\bf{r}})  &=& \mu \varepsilon _0 \omega ^2 {\bf{\Psi }}({\bf{r}})~,
\end{eqnarray}
in which the optical potential $U(\omega ,{\bf{r}})$, analogous to that in Sch\"{o}rdinger equation for electrons, is defined in the present paper by
\begin{eqnarray}
U(\omega ,{\bf{r}}) &=& \mu \varepsilon _0 \omega ^2 \chi (\omega ,{\bf{r}}) + i\omega \mu \sigma (\omega ,{\bf{r}})\nonumber\\
 &=& f(\omega ,{\bf{r}}) + i\omega g(\omega ,{\bf{r}})~,
\end{eqnarray}
where the $\chi (\omega ,{\bf{r}})$, $\sigma(\omega ,{\bf{r}})$ and $\varepsilon_0$ are the complex electric susceptibility, complex conductance and permittivity of vacuum, respectively.
It is worth noting that, when the light frequency is far away from the resonant frequencies of medium molecules, both $f(\omega,\textbf{r})$ and $g(\omega,\textbf{r})$ vary with frequency more flatly than the ordinary electric susceptibility and conductance do~\cite{Boyd2010}. Hence they are approximately frequency-independent for a narrow frequency band and can be read simply as $f(\textbf{r})$ and $g(\textbf{r})$, respectively.

It is a little more difficult to solve the Eq.(\ref{wave equation2}) due to the complexity of the optical potential and the vector nature of the electric field. An elementary approach is to expand the wave function
and optical potential in an appropriate set of generalized Fourier basis functions $\{\varphi_{lmn}(\textbf{r})\}$, let us say, a complete set in the subspace of $L^2$ functions, and to find the equations for the expansion coefficients, keeping in mind that it usually need three ``quantum numbers'' to label the basis functions in three-dimensional space. For this reason $f(\textbf{r})$, $g(\textbf{r})$ and ${\bm \Psi}(\textbf{r})$ are assumed to have following expansions,
\begin{subequations}\label{expansion1}
\begin{eqnarray}
\label{expanded permittivity}f({\bf{r}}) &=& \sum\limits_{lmn} {c^{lmn} \varphi _{lmn} ({\bf{r}})}~,\\
g({\bf{r}}) &=& \sum\limits_{lmn} {d^{lmn} \varphi _{lmn} ({\bf{r}})}~,\\
\label{expanded phi}{\bf{\Psi }}({\bf{r}}) &=& \sum\limits_{lmn} {{\bf{A}}^{lmn} \varphi _{lmn} ({\bf{r}})}~.
\end{eqnarray}
\end{subequations}
Where the scalar expansion coefficients $\{c^{lmn}\}$ and $\{d^{lmn}\}$ are, respectively, referred to as a general OSF and a general optical dissipation factor (ODF), and due to the vector nature of light each vector expansion coefficient $\textbf{A}^{lmn}$ includes three components $A_x^{lmn}$, $A_y^{lmn}$ and $A_z^{lmn}$.
By inspection of the completeness of the generalized Fourier basis functions one can rationally set that
\begin{subequations}\label{expansion2}
\begin{eqnarray}
\varphi _{hij} ({\bf{r}})\varphi _{opq} ({\bf{r}}) &=& \sum\limits_{lmn} {T_{hij,opq}^{lmn} \varphi _{lmn} ({\bf{r}})}\label{time}~,\\
\nabla \varphi _{opq} ({\bf{r}}) &=& \sum\limits_{hij} {{\bf{\Gamma }}_{opq}^{hij} \varphi _{hij} ({\bf{r}})}\label{expanded grads}~,
\end{eqnarray}
\end{subequations}
where ${T_{hij,opq}^{lmn}}$ is a scalar expansion coefficient, while ${\bf{\Gamma }}_{opq}^{hij}$  a vector expansion coefficient.
When expressions (\ref{expansion1}) for $f(\textbf{r})$, $g(\textbf{r})$ and ${\bm \Psi}(\textbf{r})$ are substituted in the Eq.(\ref{wave equation2}) and further use is made of the expansions (\ref{expansion2}), we reach a set of linear equations for expansion coefficients,
\begin{eqnarray}
\sum\limits_{opq} {\left( { - \Lambda _{opq}^{lmn}  - C_{opq}^{lmn}  - i\omega D_{opq}^{lmn}  + {\bf{\Pi }}_{opq}^{lmn}  \cdot } \right){\bf{A}}^{opq} } \nonumber\\
=\mu \varepsilon _0 \omega ^2 {\bf{A}}^{lmn}~,
\label{matrix equation 1}
\end{eqnarray}
where we make use of definitions that $\Lambda _{opq}^{lmn}  =  {{\bf{\Gamma }}_{opq}^{hij}  \cdot {\bf{\Gamma }}_{hij}^{lmn} }$, $C _{opq}^{lmn}  = {c^{hij} T_{hij,opq}^{lmn} }$, $D _{opq}^{lmn}  = {d^{hij} T_{hij,opq}^{lmn} }$ and ${\bf{\Pi }}_{opq}^{lmn}  = {{\bf{\Gamma }}_{opq}^{hij} {\bf{\Gamma }}_{hij}^{lmn} }$,
in which the Einstein summation convention with respect to repeated indices has been used.
In the following we will chose a set of eigen solutions of Helmholtz equation, denoted by $\{\varphi _{lmn} ({\bf{r}})=e^{i\textbf{k}_{lmn}\cdot\textbf{r}}\}$, which obey periodic boundary conditions within a closed square box of volume V, as the basis functions of the generalized Fourier expansions (\ref{expansion1}) and (\ref{expansion2}).
Where ${\bf{k}}_{lmn}  = l\frac{{2\pi }}{{L_x }}{\bf{\hat x}} + m\frac{{2\pi }}{{L_y }}{\bf{\hat y}} + n\frac{{2\pi }}{{L_z }}{\bf{\hat z}}$, in which ${\bf{\hat x}}$, ${\bf{\hat y}}$ and ${\bf{\hat z}}$ denote three cartesian basis vectors, $L_{x}$, $L_{y}$ and $L_{z}$  are three edge lengths of the imaginary box that should be taken large enough to contain the inhomogeneous medium we are discussing, and $l,m,n$ are integers.
This is in fact a straightforward generalization, to the case of three-dimensional disordered media, of the plain wave expansion of vector electromagnetic field developed for photonic crystals~\cite{Ho1990}.
As we shall limit our consideration to non-dissipative inhomogeneous media by removing the imaginary parts of the optical potentials,
the Eqs.(\ref{matrix equation 1}) thus become to be
\begin{widetext}
\begin{eqnarray}
\label{matrix equation 8}
\left[ {\mu \varepsilon _0 \omega ^2  + c^{000}- \left( {{\bf{k}}_{lmn}  \cdot {\bf{k}}_{lmn} } \right)} \right]{\bf{A}}^{lmn}  + {\bf{k}}_{lmn} \left( {{\bf{k}}_{lmn}  \cdot {\bf{A}}^{lmn} } \right) + \sum\limits_{opq}^{\prime} {c^{(l - o)(m - p)(n - q)} {\bf{A}}^{opq} }  &=& 0~.
\end{eqnarray}
\end{widetext}
Here putting a prime on the summation sign means that it excludes the term proportional to $c^{000}$, corresponding to the case of $o= l$, $p= m$ and $q= n$.
When $  |c|_{\max }/ |({\bf{k}}_{lmn}  \cdot {\bf{k}}_{lmn} ) - \mu \varepsilon _0 \omega ^2 - c^{000} | \ll 1$, the summation labeled by the prime can be neglected, here $|c|_{\max}=Max\{|c^{lmn}| ~| ~l\neq 0 ~or~ m\neq 0 ~or~ n\neq 0 \}$, which denotes the maximum complex modulus of that among the elements of the OSF in the summation.

A localized optical eigenmode can be deemed to be a ligh wave packet superposed by many plain-wave components exhibiting an common frequency but adjacent propagation vectors.
It is quite evident that the propagation vectors, ${\bf{k}}_{lmn}$'s, with which the related plain-wave components of a localized optical eigenmode have non-negligible amplitudes due to the coupling with each other through the OSF,
correspond to those coupled equations in Eqs.(\ref{matrix equation 8}) and obey
\begin{eqnarray}
\label{inequation}
\frac{{\left| c \right|_{\max } }}{{\left| {\left( {{\bf{k}}_{lmn}  \cdot {\bf{k}}_{lmn} } \right) - k_0^2 -c^{000}} \right|}} &>& \kappa ~,
\end{eqnarray}
with $0 < \kappa  \ll 1$, here $k_0  = \sqrt {\mu \varepsilon _0 } \omega  = 2\pi/\lambda_0$, and $\kappa$ is chosen appropriately as a parameter
determining what degree of approximation those coupled equations would be saved to.
From condition (\ref{inequation}) the uncertainty of the propagation vectors in wave vector space can be written as
\begin{eqnarray}
\Delta k_{lmn} &=& \sqrt {k_0^2  + c^{000}  + \kappa ^{ - 1} \left| c \right|_{\max } } \nonumber \\
&~& - H(k_0^2  + c^{000}  - \kappa ^{ - 1} \left| c \right|_{\max } )\nonumber \\
&~& ~~~\times\sqrt {k_0^2  + c^{000}  - \kappa ^{ - 1} \left| c \right|_{\max } }~,
\end{eqnarray}
where $H(x)$ is the Heaviside step function.
The appearance of a localized light wave packet with feature size of $\xi$ requires, according to the uncertainty principle, that
$\Delta k_{lmn} \xi \gtrsim 4\pi $, in which the $\xi$ is also a typical coherence length of light field because of the stationary space correlation of the electromagnetic vibration within such a wave packet.
In addition we define the effective wavelengths of the light waves in an inhomogeneous medium as the vacuum wavelengths divided by the root mean square refractive index, i.e., $\lambda _{eff}  =\lambda_0 / \sqrt {\left\langle {n^2 ({\bf{r}})} \right\rangle }$, in which $\left\langle {n^2 ({\bf{r}})} \right\rangle  = \left\langle {\chi({\bf{r}})} \right\rangle + 1 = c^{000} / k_0^2 +1$, and the angle brackets represents the configurational averaging of the function included in it.
It seems natural to work with the ratio of the effective wavelength to the feature size of the localized light wave packet as a convenient measurement for the degree of localization. Therefore we rewrite, for any of eigen optical modes represented by a single wave packet, the uncertainty principle in the form
\begin{eqnarray}
\label{criterion}
G \equiv \frac{{\Delta k_{lmn} }}{{2\sqrt {k_0^2  + c^{000} } }} \gtrsim \frac{{\lambda _{eff} }}{\xi }~,
\end{eqnarray}
where the dimensionless ratio, $ G $, is introduced as a localization intensity parameter which closely relates to the OSF.

Based on the consideration that the feature size of the wave packet is of the same order of magnitude with the elastic mean free path entering the standard Ioffe-Regel condition~\cite{Ioffe1960},
the approximate criterion for strong localization of light, ${{\lambda _{eff} } \mathord{\left/ {\vphantom {{\lambda _{eff} } \xi }} \right.
\kern-\nulldelimiterspace} \xi }\gtrapprox 1 $, is likely to be satisfied only when $G$ is large enough to nearing unit,
otherwise the light field would be in extended state.
In fact, the $G$ exhibits a natural singular value $1/\sqrt{2}$ corresponding to the discontinuity point of the included Heaviside step function. When $G$ takes positive values far below $1/\sqrt{2}$ the wave vectors, $\textbf{k}_{lmn}$'s, fulfilling the inequality (\ref{inequation}), are distribute in a thin spherical shell centered at the origin of $k$ space, on which optical eigenmodes are easy to be developed as modulated carrier waves with large-size envelopes. As the $G$ rises above $1/\sqrt{2}$ the thickness of this spherical shell increases and, meantime, its inner surface shrinks eventually to the origin so that there is an abruptly increasing in the uncertainty of the wave vectors, with which optical eigenmodes will have chances of forming small-size standing-wave packets.
Thus it is our hypothesis that $G > 1/\sqrt{2}$ may provide the key to the observation of strong localization and pattern transition of photons in certain dissipative inhomogeneous media.

However what is amazing is that $G \sim \lambda_{eff} / \xi $ is very similar to the Ginzburg-Landau parameter in the phenomenological theory of superconductivity~\cite{Lifshitz1980}, which is used to distinguish two typical classes of superconductors, i.e., Type I superconductors characterized by the extended states of superconducting electrons are those with Ginzburg-Landau parameter less than $1/\sqrt{2}$, and Type II superconductors supporting the mixed states of superconducting electrons those with Ginzburg-Landau parameter large than $1/\sqrt{2}$~\cite{Abrikosov1957}.
Furthermore, a detailed analysis shows that there may be a photonic phase supporting the coexistence of optical modes with different localization degrees although $G>1/\sqrt{2}$ \cite{SupplementalMaterial}. One amplified radiation version of the coexistent optical modes, achieved by introducing a gain mechanism, would be probed through the spectrally resolved speckle technique~\cite{Cao2002}.  This argument has been verified by the recent experiments investigating the random laser modes in the nanocrystalline ZnO powder~\cite{Johannes2009}.
Generally speaking, the real light field is a linear superposition of various optical eigenmodes, whose proportions depend on the boundary condition. Specially in the thermodynamic equilibrium of a finite temperature, photons obey the Bose-Einstein distribution and occupy a variety of eigenmodes.
In this regard we believe that when $G < 1/\sqrt{2}$ there is an ES phase, characterized by the exclusively extended states of all optical eigenmodes in the narrow frequency window matching with a frequency-independent optical potential, while $G > 1/\sqrt{2}$ indicates a MS phase, in which some eigenmodes exist in extended states and the others belong to localized states. The boundary is specified by $G \sim 1/\sqrt{2}$ where a photonic phase transition will occur: the onset of a strong localization transition of some photons.

\begin{figure}[h]
\begin{center}
\includegraphics[width=8.0cm]{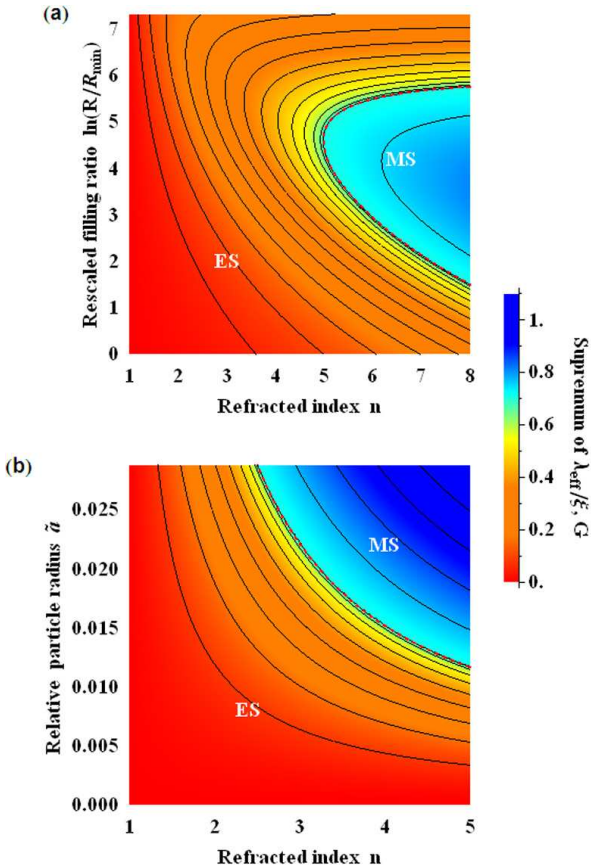}
\end{center}

\caption{Colour-coded graphs showing the localization intensity parameter, $G$, as a function of the refractive index $n$,
the rescaled filling ratio $\text{ln}(R/R_{min})$, and the relative particle radius $\tilde{a}$, for the adjustable parameters
taking values of $\kappa=0.005$ and $N_{min}=100$. Slices (a) and (b) correspond to the cross sections at $\tilde{a}=0.01$(a)
and $R=0.01$(b), respectively. Extended-state (ES) phase exists in the regions of $G<1/\sqrt{2}$, as dominated mainly by the orange-red,
and the mixed-state (MS) phase is found on the regions of $G>1/\sqrt{2}$, as controlled almost by blue.
The Black solid lines are contours of $G$, and the red dashed line corresponds to the critical point where $G=1/\sqrt{2}$,
delineating the photonic phase transition.}
\label{sketch map}
\end{figure}

Further utilization of the uncertainty inequality (\ref{criterion}) requires the evaluation of the OSF of a concrete medium.
To avoid the complication of calculation we will only consider a disordered, non-dissipative medium that consist of $N$ perfectly spherical scattering particles with both an uniform radius $a$ and an identical reflective index $n$, which are immersed randomly in a background medium of dielectric constant $\varepsilon_0$.
In this case we define
$R = {{4\pi a^3 N} \mathord{\left/
 {\vphantom {{4\pi a^3 N} {\left( {3V} \right)}}} \right.
 \kern-\nulldelimiterspace} {\left( {3V} \right)}} = {{4\pi \tilde a^3 N} \mathord{\left/
 {\vphantom {{4\pi \tilde a^3 N} 3}} \right.
 \kern-\nulldelimiterspace} 3}$ as the filling ratio of the scattering particles, in which $V=L_{x}L_{y}L_{z}$ is the volume of the closed square box over which the Fourier expansion is defined, and $\tilde a = aV^{{{ - 1} \mathord{\left/
 {\vphantom {{ - 1} 3}} \right. \kern-\nulldelimiterspace} 3}}$ a dimensionless relative radius.
The $G$ can be written as a variety of functions of the actual optical and geometric parameters of this disordered medium \cite{SupplementalMaterial},
\begin{eqnarray}
\label{LIP}
G &=& \frac{1}{2}\left( {\sqrt {1 + \Omega }  - H(1 - \Omega )\sqrt {1 - \Omega } } \right)~,
\end{eqnarray}
in which $\Omega$ have some equivalent explicit expressions such as
\begin{subequations}\label{state function}
\begin{eqnarray}
\Omega (n,R,N) &=& \frac{{R\left( {n^2  - 1} \right)}}{{1 + R\left( {n^2  - 1} \right)}}\frac{1}{{\kappa \sqrt N }}~,
\end{eqnarray}
${\rm{for}} ~ R \leq R_{\max} ~ {\rm{and}}~ N \geq N_{\min}$, and
\begin{eqnarray}
\Omega (n,R,\tilde a) &=& \frac{{R\left( {n^2  - 1} \right)}}{{1 + R\left( {n^2  - 1} \right)}}\frac{1}{\kappa }\sqrt {\frac{{3R}}{{4\pi \tilde a^3 }}}~,
\end{eqnarray}
${\rm{for}} ~ R_{\min }  \le R \le R_{\max } ~
 {\rm{and}} ~ \tilde a \le \left[ {{{3R} \mathord{\left/ {\vphantom {{3R} {\left( {4\pi N_{\min } } \right)}}} \right.
\kern-\nulldelimiterspace} {\left( {4\pi N_{\min } } \right)}}} \right]^{\frac{1}{3}}$.
\end{subequations}
Where ${{R_{\min }  = N_{\min } 4\pi \tilde a^3 } \mathord{\left/ {\vphantom {{R_{\min }  = N_{\min } 4\pi \tilde a^3 } 3}} \right.
\kern-\nulldelimiterspace} 3}$ is the minimum filling ratio for a fixed relative radius of $\tilde a$,
$N_{\min} \gg 1$ is the corresponding minimum number of the filling particles,
and ${R_{\max}}$ is the maximum filling ratio determined by the random close packing of the spherical particles, for which different types of experiments have suggested an value of about $0.637$ \cite{zallen1983}.
It is worth noting here that $N_{\min}$ and $\kappa$, as they only associate with the approximation procedure, can serve as adjustable parameters when fitting experimental results.
In Fig.\ref{sketch map} we plot the contour and density graphs of the localization intensity parameter $G$ following from the expressions (\ref{LIP}) and (\ref{state function}). It unequivocally demonstrates a photonic phase transition between the ES phase and the MS phase by altering the optical and geometric parameters of the disordered medium \cite {SupplementalMaterial}.

In conclusion, In this paper we have suggested a generalized mode-expansion scheme for the vector electromagnetic waves propagating in 3D disordered media. Based on the Fourier analysis we have provided an efficient way to estimate the coherence length $\xi$ of the optical eigenmodes of an inhomogeneous medium, and proposed a necessary condition, $G > 1/\sqrt{2}$, for the strong localization of light.
Via changing the optical and geometric parameters, such as the refractive index, the dimension and the density of randomly distributed dielectric spheres, we can achieve the photonic phase transition from the ES phase to the MS phase or the converse transition.
We hope that this method can be generalized to anisotropic polarized inhomogeneous media, nonlinear inhomogeneous media as well as
the similar case involving other wave phenomena. It remains to be see wether concepts from superconductivity theory and
classical fermionic systems will prove fruitful in the systems of non-free bosons maintained by disordered media.

We wish to acknowledge the supports of the NSF Grant No.10874016 and the Program for Hunan Provincial Optical Key Discipline of China.

\end{document}